\documentclass{RAA}
\usepackage{graphicx,times}
\usepackage{natbib}
\bibpunct[, ]{(}{)}{;}{a}{}{,}

\def\bear{\begin{eqnarray}}
\def\ear{\end{eqnarray}}

\newcommand{\Ds}{{D_{\rm s}}}
\newcommand{\Dd}{{D_{\rm d}}}
\newcommand{\Dds}{{D_{\rm ds}}}
\newcommand{\rE}{{r_{\rm E}}}
\newcommand{\tE}{{t_{\rm E}}}
\newcommand{\vt}{{v_{\rm t}}}
\newcommand{\murel}{{\mu_{\rm rel}}}
\newcommand{\thetaE}{{\theta_{\rm E}}}
\newcommand{\ys}{{y_{\rm s}}}
\newcommand{\xs}{{x_{\rm s}}}
\newcommand{\rs}{{r_{\rm s}}}
\newcommand{\zs}{{z_{\rm s}}}
\newcommand{\Fs}{{{\cal F}_{\rm s}}}
\newcommand{\ri}{{r}}
\def\yr{\,{\rm yr}}
\def\day{\,{\rm day}}

\def\kpc{\,{\rm kpc}}
\def\kms{\,{\rm km\,s^{-1}}}

\def\AU{\,{\rm AU}}
\def\mas{\,{\rm mas}}

\newcommand{\beq}{\begin{equation}}
\newcommand{\eeq}{\end{equation}}

\newcommand{\sect}[1]{Sec.\,#1}

\begin{document}
\title{Astrophysical Applications of Gravitational Microlensing}
\volnopage{ {\bf 2012} Vol.\ {\bf 12} No. {\bf 1},~ 000 --- 010}
\setcounter{page}{000}

\author{Shude Mao\inst{1,2}}
\institute{National Astronomical Observatories, Beijing 100012, China \\
\and
Jodrell Bank Centre for Astrophysics, Alan Turing Building, University
of Manchester, Manchester M13 9PL, UK;
{\it smao@nao.cas.cn; shude.mao@manchester.ac.uk} \\
\vs \no
   {\small Received 2012 June 30; accepted 2012 July 30}
}

\abstract{
Since the first discovery of microlensing events nearly two decades ago,
gravitational microlensing has accumulated tens of TBytes of data and 
developed into a powerful astrophysical technique with diverse applications. The review starts
with a theoretical overview of the field and then proceeds to discuss the scientific
highlights.
(1) Microlensing observations toward the Magellanic Clouds rule out the Milky Way halo being dominated by MAssive Compact Halo Objects (MACHOs). This confirms most dark matter is non-baryonic, consistent with other observations.
(2) Microlensing has discovered about 20 extrasolar planets (16 published),
including the first two Jupiter-Saturn like systems and the only
``cold Neptunes" yet detected. They probe a
different part of the parameter space and will likely provide the most stringent test of core accretion theory of planet formation.
(3) Microlensing provides a unique way to measure the mass of isolated
stars, including brown dwarfs to normal stars. Half a dozen or so stellar mass
black hole candidates have also been proposed. 
(4) High-resolution, target-of-opportunity
spectra of highly-magnified dwarf stars provide intriguing ``age''
determinations which may either hint at enhanced helium enrichment or
unusual bulge formation theories.
(5) Microlensing also measured limb-darkening
profiles for close to ten giant stars, which challenges stellar atmosphere
models.
(6) Data from surveys also provide strong
constraints on the geometry and kinematics of the Milky Way bar (through proper
motions); the latter indicates predictions from current models appear to
be too anisotropic compared with observations.
The future of microlensing is bright given the new
capabilities of current surveys and forthcoming new telescope networks
from the ground and from space. Some open issues in the field are
identified and briefly discussed.
}


\authorrunning{Mao}
\titlerunning{Astrophysical Applications of Gravitational Microlensing}
\vspace{3mm} \no{\sf INVITED REVIEWS}
\maketitle

\section{Introduction} \label{sec:intro}

Gravitational microlensing in the local group refers to the temporal
brightening of a background star due to intervening objects.
\cite{Ein36} first studied (micro)lensing by a single
star, and concluded that ``there is no great chance of observing this
phenomenon.'' Although there were some works in intervening years by \cite{Ref64}
and \cite{Lie64}, the field was revitalized by \cite{Pac86}
who proposed it as a method to detect MAssive Compact Halo Objects (MACHOs) in the Galactic halo.

From observations of microwave background radiation and
nucleosynthesis (see, e.g. \citealt{Kom11, Ste07}), it is clear that most of the dark matter must be
non-baryonic, and so the original goal of microlensing is now obsolete.
Nevertheless, microlensing has developed into a powerful
technique with diverse applications in
astrophysics, including constraints on MACHOs, the study of the structure of the Milky Way,
stellar atmospheres and the detection of extrasolar planets and
stellar-mass black hole candidates. Since the first discoveries of microlensing events in 1993 \citep{Alc93,  Uda93}, the field has made  enormous progress in the last two decades. A
number of reviews have been written on this topic (e.g. \citealt{Pac96,
  Mao01, Eva03, Wam06}), with the
most recent highlights given in \cite{Mao08a}, \cite{Gou09a} and \cite{Gau10}. The
readers will also greatly benefit  from two recent, comprehensive
conference proceedings: the Manchester Microlensing
Conference\footnote{http://pos.sissa.it/cgi-bin/reader/conf.cgi?confid=54}
and the 2011 Sagan Exoplanet Summer Workshop: Exploring Exoplanets with
Microlensing\footnote{http://nexsci.caltech.edu/workshop/2011/}. The
workshop materials contain not only recent scientific highlights
but also hands-on exercises for data reduction and modelling.

The structure of this review is as follows. \sect\ref{sec:basics}
introduces the basics of gravitational microlensing, which reproduces
\cite{Mao08b} in a slightly modified form; \sect\ref{sec:apps} builds on the
introduction and discusses the applications of gravitational
microlensing. We finish this review with an outlook
for the field in \sect\ref{sec:future}. Due to the rapid expansion of the field,
it is unavoidable that the reference list is incomplete (and somewhat biased).

\section{Basics of Gravitational Microlensing} \label{sec:basics}

\subsection{What is gravitational microlensing?}

\begin{figure}[!ht]
\centering
\vspace{-3mm}
\includegraphics[width=.45\textwidth]{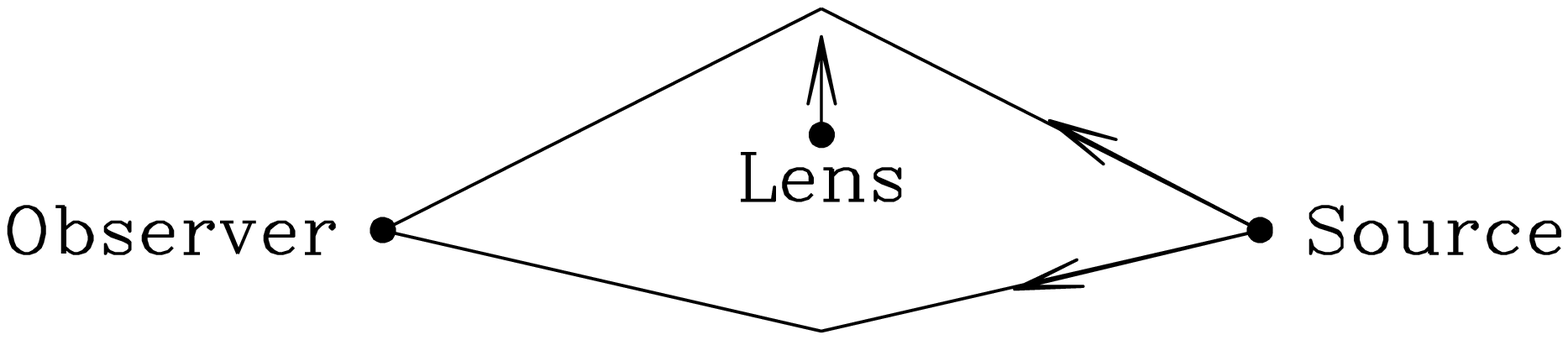} \hfil
\includegraphics[width=.45\textwidth]{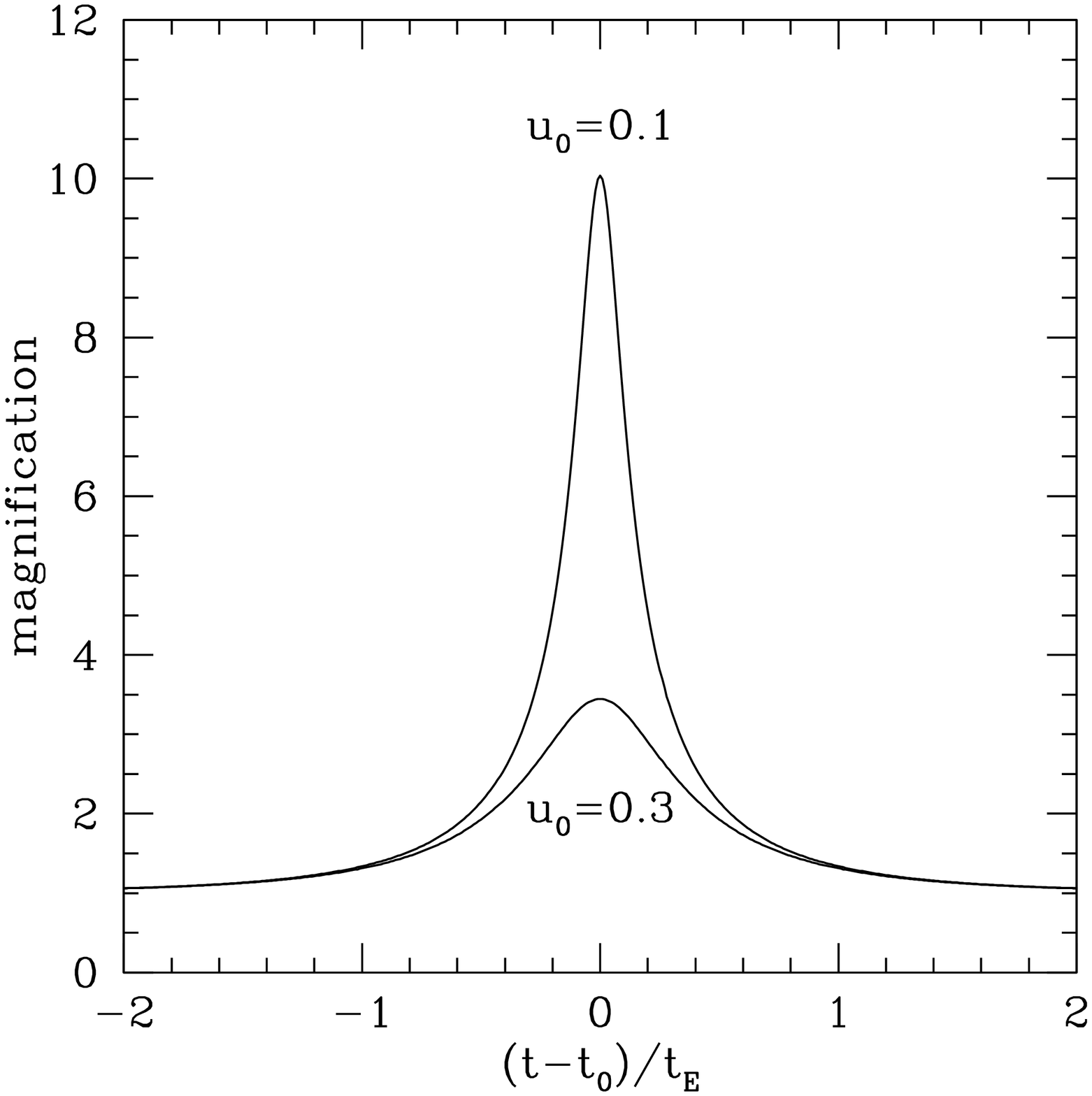}
\caption{The left panel shows a side-on view of the geometry of microlensing
  where a lens moves across the line of sight toward a background source. The
  right panel shows two light curves corresponding to two dimensionless impact
  parameters, $u_0=0.1$ and 0.3. The time on the horizontal axis is centered on the
peak time $t_0$ and is normalized to the Einstein radius crossing time
$\tE$. The lower the value of $u_0$, the higher the peak
magnification. For the definitions of $u_0$ and $\tE$ see \sect\protect\ref{sec:standard}.}
\label{fig:lc}
\end{figure}

According to General Relativity, the light from a background source is deflected, distorted and
(de)magnified by intervening objects along the line of sight. If the
lens, source and observer are sufficiently well aligned, then strong gravitational
lensing can occur. Depending on the lensing object, strong gravitational
lensing can be divided into three areas: microlensing by stars,
multiple-images by galaxies, and giant arcs and large-separation lenses
by clusters of galaxies. For microlensing, the lensing object is a stellar-mass compact object (e.g. normal
stars, brown dwarfs or stellar remnants [white dwarfs, neutron stars and
  black holes]); the image splitting in this case is
usually too small (of the order of a milli-arcsecond in the local group) to be resolved by
ground-based telescopes, thus we can only observe the change in magnification as a function of time.

The left panel in Fig.  \ref{fig:lc} illustrates the
geometry of microlensing. A stellar-mass lens moves across the line of sight toward a
background star. As the lens moves closer to the line of sight, its gravitational
focusing increases, and the background star becomes brighter. As the source moves
away, the star falls back to its baseline brightness. If the motions of
the lens, the observer and the source can be approximately taken as
linear, then the light curve is symmetric. Since the lensing probability
for microlensing in the local group is of the order of $10^{-6}$
(see \sect\ref{sec:tau}), the microlensing variability usually should not
repeat. Since photons of different wavelengths follow the same propagation path
(geodesics), the light curve (for a point source) should
not depend on the color. The characteristic symmetric shape,
non-repeatability, and achromaticity can be used as criteria to separate microlensing
from other types of variable stars (exceptions to these rules will be
discussed in \sect\ref{sec:exotic}).

To derive the characteristic light curve shape shown in the right panel
of Fig. \ref{fig:lc}, we must look closely at the lens equation, and the resulting image positions and magnifications for a point source.

\subsection{Lens equation}

\begin{figure}[!ht]
\begin{center}
\includegraphics[width=0.9\textwidth]{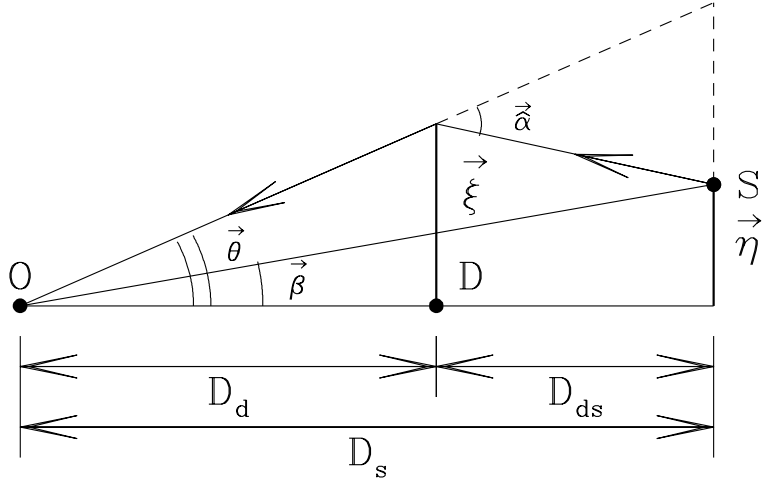}
\end{center}
\vspace{-6.5cm}
\caption{Illustration of various distances and angles in the lens 
equation (eqs. \protect\ref{eq:lensPhysical} and \protect\ref{eq:vectorlens}).
}
\label{fig:lenseq}
\end{figure}

The lens equation is straightforward to derive. From Figure 
\ref{fig:lenseq} of the lensing
configuration, simple geometry yields
\beq
\label{eq:lensPhysical}
\vec{\eta} + \Dds \hat{\vec{\alpha}} = \vec{\xi} \cdot \frac{\Ds}{\Dd},
\eeq
where $\Dd$, $\Ds$ and $\Dds$ are the distance to the lens (deflector), distance to the
source and distance between the lens (deflector) and the source,
$\vec{\eta}$ is the source position (distance perpendicular to the line
connecting the observer and the lens), $\vec{\xi}$ is the image position,
and $\hat{\vec{\alpha}}$ is the deflection angle. For gravitational microlensing in
the local group, Euclidean geometry applies and $\Dds=\Ds-\Dd$. Mathematically, the lens equation provides a mapping from the source plane to the lens plane. The mapping is not necessarily one-to-one.

Dividing both sides of Equation (\ref{eq:lensPhysical}) by $\Ds$, we obtain
the lens equation in angles
\beq
\vec{\beta} + \vec{\alpha} = \vec{\theta},
\label{eq:vectorlens}
\eeq
where $\vec{\beta}=\vec{\eta}/\Ds$, $\vec{\theta}=\vec{\xi}/\Dd$, and
$\vec{\alpha} = \hat{\vec{\alpha}} \times \Dds/\Ds$ is the scaled
deflection angle. These angles are illustrated in Fig. \ref{fig:lenseq}.

For an axis-symmetric mass distribution, due to symmetry, the source, observer and image positions
must lie in the same plane, and so we can drop the vector sign, and
obtain a scalar lens equation:
\beq
{\beta} + {\alpha} = {\theta}.
\eeq

\subsection{Image positions for a point lens}

For a point lens at the origin, the deflection angle is given by
\beq
\hat{\vec{\alpha}} = \frac{4 G M}{c^2} \frac{1}{\xi^2} \vec{\xi},
\eeq
and the value of the scaled deflection angle is
\beq
\alpha = \frac{\Dds}{\Ds} |\hat{\vec{\alpha}}|=
\frac{\Dds}{\Ds} \frac{4 G M}{c^2 \Dd \theta} \equiv \frac{\thetaE^2}{\theta}, ~~~ \xi=\Dd \theta,
\eeq
where we have defined the angular Einstein radius as
\beq
\thetaE = \frac{\rE}{\Dd} \approx
0.55 \mas \sqrt{\frac{1-\Dd/\Ds}{\Dd/Ds}} \left(\frac{\Ds}{8 \kpc}\right)^{-1/2}\left(\frac{M}{0.3 M_\odot}\right)^{1/2}.
\label{eq:thetaE}
\eeq
The lens equation for a point lens in angles is therefore
\beq
\beta + \frac{\thetaE^2}{\theta} = \theta.
\eeq
We can further simplify by normalizing all the angles by $\thetaE$,
$\rs \equiv \beta/\thetaE$, and $\ri \equiv \theta/\thetaE$, then the above
equation becomes
\beq
\rs  + \frac{1}{\ri} = \ri,
\label{eq:rs}
\eeq
where $\rs$ is the source position and not to be confused with the size of
the star, which we denote as $r_\star$.

For the special case when the lens, source and observer are perfectly
aligned ($\rs=0$), due to axis-symmetry along the line of sight, the
images form a ring (``Einstein'' ring) with its angular size given by Equation (\ref{eq:thetaE}).

For any other source position $\rs \ne 0$, there are always two images, whose
positions are given by
\beq
\ri_{\pm} = \frac{\rs \pm \sqrt{\rs^2+4}}{2}.
\label{eq:image0}
\eeq
The `+' image is outside the Einstein radius ($\ri_{+} \ge 1$) on the same
side as the source, while the `$-$' image is on the opposite side and
inside the Einstein radius ($\ri_{-}<0$ and $|\ri_{-}|<1$). The angular separation between
the two images is
\beq
\Delta \theta = \thetaE (\ri_{+} - \ri_{-}) = \thetaE \sqrt{4+\rs^2}.
\eeq
The image separation is of the same order as the angular Einstein diameter when
$\rs \ls 1$, and thus will be in general too small to be observable given the typical
seeing from the ground ($\sim$ one arcsecond); we can only observe
lensing effects through magnification. One exception may be the VLT interferometer (VLTI)
which can potentially resolve the two images. This may be important for
discovering stellar-mass black holes since they have larger image
separations due to their larger masses than typical lenses with mass
$\sim 0.3 M_\odot$ (\citealt{Del01, RM06}).

The physical size of the Einstein radius in the lens plane is given by
\begin{equation}
\rE = \Dd\thetaE = \sqrt{ \frac{4 GM}{c^2} \frac{\Dd \Dds}{\Ds}}
\approx 2.2 \AU \sqrt{4 \times \frac{\Dd}{\Ds}\left(1-\frac{\Dd}{\Ds}\right)} \left(\frac{\Ds}{8 \kpc}\right)^{1/2}  \left(\frac{M}{0.3 M_\odot}\right)^{1/2}. 
\label{eq:rE}
\end{equation}
So the size of the Einstein ring is roughly the scale of the solar
system, which is a coincidence that helps the discovery of extrasolar
planets around lenses.

\subsection{Image magnifications}

Since gravitational lensing conserves surface
brightness\footnote{Imagine looking at a piece of white paper with a
  magnifying glass, the area is magnified, but the brightness of the
  paper is the same.}, the 
magnification of an image is simply given by the ratio of the image area
and source area, which in general is given by the
 determinant of the Jacobian in the lens mapping (see \sect\ref{sec:binary}).

\begin{figure}[!t]
\vspace{-1cm}
\begin{center}
\includegraphics[width=0.7\textwidth]{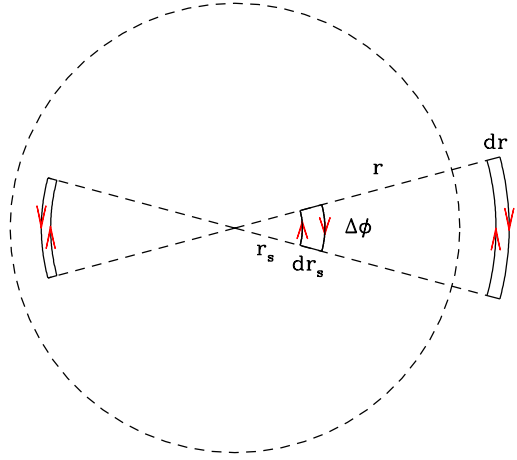}
\end{center}
\vspace{-2.5cm}
\caption{Images of a thin annulus from $\rs$ to $\rs+d\rs$ by a point
lens on the plane of the sky. The dashed line is the Einstein
ring. $\Delta\phi$ is the angle subtended by the thin annulus. Notice
that the positive image (outside the Einstein ring) is clockwise while
the negative image (inside the Einstein ring) is counter-clockwise.
}
\label{fig:mag}
\end{figure}

Here we attempt a more intuitive derivation. For a very small source, we can consider a thin source
annulus with angle $\Delta \phi$ (see Fig. \ref{fig:mag}). For a point lens, this
thin annulus will be mapped into two annuli, one inside the Einstein ring and one outside.

The area of the source annulus is given by the product of the radial
width and the tangential length $d\rs \times \rs\Delta\phi$. Similarly, 
each image area is $d\ri \times \ri\Delta\phi$, and the
magnification is given by

\beq
\mu = \frac{ d\ri \times \ri\Delta\phi}{d\rs \times \rs\Delta\phi} 
= \frac{\ri}{\rs} \frac{d\ri}{d\rs}.
\label{eq:image}
\eeq
For two images given in Equation (\ref{eq:image0}), one finds
\beq
\mu_{+} = \frac{(\rs+\sqrt{\rs^2+4})^2}{4 \rs \sqrt{\rs^2+4}}, ~~
\mu_{-} = -\frac{(\rs-\sqrt{\rs^2+4})^2}{4 \rs \sqrt{\rs^2+4}}.
\eeq
The magnification of the `+' image is positive, while the `$-$' image is
negative. The former image is said to have positive parity while the
latter is negative (for the geometric meaning, see Fig. \ref{fig:mag}).
The total magnification is given by
\beq
\mu = |\mu_{+}| +|\mu_{-}| = \frac{\rs^2+2}{\rs\sqrt{\rs^2+4}},
\label{eq:mu}
\eeq
and the difference is identically equal to unity
\beq
|\mu_{+}|-|\mu_{-}|=1.
\eeq
We make some remarks about the total magnification and image separations:
\begin{enumerate}
\item When $\rs=1$, $\mu=3/\sqrt{5}\approx 1.342$,  corresponding to about 0.319
  magnitude. Such a difference is easily observable (For bright
stars, the accuracy of photometry can reach a few milli-magnitudes.), and so the area
  occupied by the Einstein ring is usually taken as the lensing ``cross-section.''
\item When $\rs \rightarrow \infty$, $|\mu_{+}/\mu_{-}| \rightarrow
  \rs^{4}$, $\mu \rightarrow 1+2 \rs^{-4}$. The angular image
  separation is given by $\Delta\theta = (\rs+2 \rs^{-1})\thetaE$.
\item High magnification events occur when $\rs \rightarrow 0$. 
The asymptotic behaviors are $\mu \rightarrow \rs^{-1}(1+3 \rs^2/8)$, $\Delta\theta
\rightarrow (2+\rs^2/4)\thetaE$, and $dr/d\rs \rightarrow 1/2$. The last
relation implies that, at high magnification, the image is compressed by a
factor of two in the radial direction (see Fig. \ref{fig:mag}).
\end{enumerate}

\subsection{Light curve and microlensing degeneracy}

Given a source trajectory, we can easily describe the standard light curve with
a few parameters which suffers from the microlensing degeneracy.

\subsubsection{Standard light curve \label{sec:standard}}

\begin{figure}[!t]
\vspace{-1.5cm}
\begin{center}
\includegraphics[width=0.7\textwidth]{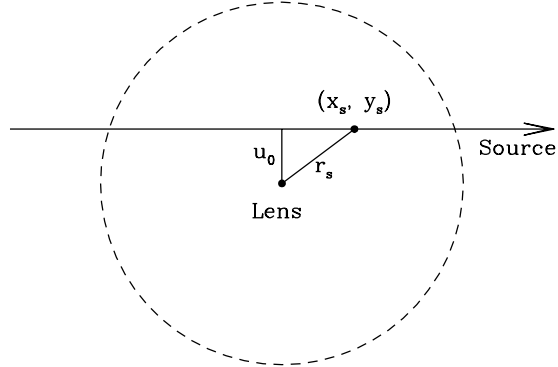}
\end{center}
\vspace{-3cm}
\caption{Illustration of the lens position and source trajectory. The
  dimensionless impact parameter is $u_0$,
$(\xs, \ys)$ is the dimensionless source position along the trajectory,
and $\rs$ is the distance between the lens and source.}
\label{fig:traj}
\end{figure}

For convenience, we put the lens at the origin, and let the source move across the line
of sight along the $x$-axis (see Fig. \ref{fig:traj}). The impact parameter
in units of the Einstein radius is labeled as $u_0$. For convenience,
we define the Einstein radius crossing time (or `timescale') as
\beq
\tE = \frac{\rE}{\vt} = \frac{\thetaE}{\murel}, ~~
\thetaE=\frac{\rE}{\Dd}, ~~ \murel \equiv \frac{\vt}{\Dd},
\label{eq:tE}
\eeq
where $\vt$ is the transverse velocity (on the lens plane) and $\murel$ is the relative
lens-source proper motion. Substituting the expression for the Einstein
radius into Equation (\ref{eq:rE}), we find that
\beq
\tE \approx 19\,\day \sqrt{4\times \frac{\Dd}{\Ds}\left(1-\frac{\Dd}{\Ds}\right)} \left(\frac{\Ds}{8 \kpc}\right)^{1/2} 
\left(\frac{M}{0.3M_\odot}\right)^{1/2}
\left(\frac{\vt}{200\kms}\right)^{-1}.
\eeq
If the closest approach is achieved at time $t=t_0$, then the
dimensionless coordinates are $\xs = (t-t_0)/\tE$, $\ys = u_0$ and the
magnification as a function of time is given by
\beq
\mu(t) = \frac{\rs^2(t)+2}{\rs(t)\sqrt{\rs^2(t)+4}}, ~~~ \rs(t) =
\sqrt{u_0^2 + \left(\frac{t-t_0}{\tE}\right)^2}.
\label{eq:standard}
\eeq
Two examples of light curves are shown in the right panel of Fig. \ref{fig:lc} for $u_0=0.1$ and 0.3.

To model an observed light curve, three parameters are present in Equation (\ref{eq:standard}): $t_0$,
$\tE$, and $u_0$. In practice, we need two additional parameters, $m_0$, the
baseline magnitude, and $\Fs$, a blending parameter. $\Fs$ 
characterizes the fraction of light contributed by the lensed source; in
crowded stellar fields, each observed `star' may be a composite of the
lensed star, other unrelated stars within the seeing disk and the lens
if it is luminous (\citealt{Alc01b, Koz07}). Blending will lower the observed magnification and in general $\Fs$ depends on the
wavelength, and so each filter requires a separate $\Fs$ parameter.

Unfortunately, we can see from Equation (\ref{eq:standard}) that
there is only one physical parameter ($\tE$) in the model that relates to the lens' properties.
Since $\tE$ depends on the lens mass, distances to the lens and source, and
the transverse velocity $\murel$, from an observed light
curve well fitted by the standard model, one cannot uniquely infer the
lens distance and mass; this is the so-called microlensing
degeneracy. However, given a lens mass function and some kinematic model of the
Milky Way, we can infer the lens mass statistically.

\subsection{Non-standard light curves \label{sec:exotic}}

The standard model assumes the lensed source is point-like, both the
lens and source are single and all the motions are linear.
The majority ($\sim 90\%$) of microlensing events are well described by
this simple model. However, about 10\% of the light curves are
non-standard (exotic), due to  the breakdown of one (or more) of the assumptions.
We briefly list these possibilities below.
These non-standard microlensing events allow us to derive extra constraints, and partially
lift the microlensing degeneracy. Because of this, they play a role far
greater than their numbers suggest.
\begin{enumerate}
\item[(1)] {\it binary lens events:} The lens may be in a binary or even a multiple system
(\citealt{MP91}). The light curves for a binary or multiple lensing system
can be much more diverse (see \sect\ref{sec:binary}). They offer an exciting
way to discover extrasolar planets (\citealt{MP91, GL92, BR96, GS98,
  Rat02}, for more see \sect\ref{sec:planets}).
\item[(2)] {\it binary source events:} The source is in a binary system. In this case, the light curve will
  be a simple, linear superposition of the two sources (when the orbital
  motion can be
  neglected, see \citealt{GH92}).
\item[(3)] {\it finite source size events:} The finite size of the lensed star cannot be neglected. This
occurs when the impact parameter $u_0$ is comparable to the stellar
radius normalized to the Einstein radius, $u_0 \sim r_\star/\rE$. In
this case, the light curve is significantly modified by the finite
source size effect (\citealt{WM94, Gou94}). The finite source size effect
is most important for high magnification events.
\item[(4)] {\it parallax/``xallarap'' events:} The standard light curve assumes all the motions are
linear. However, the source and/or the lens may be in a binary system,
furthermore, the Earth revolves around the Sun. All these motions induce
accelerations. The effect due to Earth's motion around the Sun is
usually called ``parallax'' (e.g. \citealt{Gou92, Smi02, Poi05}), while that due to binary motion in the
source plane is called ``xallarap'' (``parallax'' spelt backwards,
\citealt{Ben98, Alc01a}). Parallax or ``xallarap'' events usually have long
timescales. For a typical microlensing event with timescale $\tE \sim
20\day$, the parallax effect due to the Earth rotation around the Sun is often undetectable (unless
the photometric accuracy of the light curve is very high or the lens is
very close).
\item[(5)] {\it Repeating events:}
  Microlensing can ``repeat'', in particular if the lens is a wide
  binary (\citealt{Dis96}) or the source is a wide binary. In such cases, microlensing may
  manifest as two well-separated peaks, i.e., as a ``repeating'' event. A
  few percent of microlensing events are predicted to repeat, consistent
  with the observations (\citealt{Sko09}).
\end{enumerate}
Notice that several violations may occur at the same time, which in some
cases allow the microlensing degeneracy to be completely removed
(e.g. \citealt{An02, Don09b, Gau08}).

In particular, when both finite source size and parallax effects are seen, the lens mass can be
determined uniquely \citep{Gou92}. Microlensing
parallax measures the projected Einstein radius in the observer plane
(in units of the Earth-Sun separation, AU):
$\tilde{r}_{\rm E}=\rE \times \Ds/\Dds$, or equivalently, the parallax
$\pi_{\rm E}=1\,{\rm AU}/\tilde{r}_{\rm E}$.
The finite source size effect, on the other hand,
measures the ratio of the angular stellar radius
$\theta_\star$ to the angular Einstein radius $\thetaE$. Since the angular size of the lensed star can
be measured from its color and apparent magnitude \citep{Yoo04}, we can derive
the angular Einstein radius $\thetaE$. It is straightforward to
combine Eqs. (\ref{eq:rE}) and (\ref{eq:thetaE}) to obtain
\beq
M={c^2 \over 4G}\tilde{r}_{\rm E}\,\thetaE =
{c^2 \over 4G} {\thetaE \over \pi_{\rm E}}.
\label{eq:mass}
\eeq
Notice that the determination is independent of all the
distances. 

Equation (\ref{eq:mass}) is especially transparent in the 
natural formalism advocated by \cite{Gou00}, which provides a way to
connect quantities measurable in microlensing ($\pi_{\rm E},
\thetaE, \tE$) with other physical quantities:
\beq
\pi_{\rm rel} = \pi_{\rm E} \thetaE,~~
\mu_{\rm rel} = {\thetaE \over \tE},
\eeq
where $\pi_E=1/\tilde{r}_{\rm E}$, $\pi_{\rm rel}=1/\Dd - 1/\Ds$ is the lens-source relative
parallax and $\mu_{\rm rel}$ is the relative proper motion (also used in
equation \ref{eq:tE}). For example,
if the lens and distance can be measured (using other means), then $\pi_{\rm rel}$ can be
determined. In addition, if the lens and source motions can be
measured, then we can determine the relative proper motions ($\mu_{\rm rel}$).

\subsection{$N$-point lens gravitational microlensing \label{sec:binary}}

It is straightforward to derive the (dimensionless) lens equation for $N$-point lenses. We can first cast
Equation (\ref{eq:rs}) in vector form and then rearrange
\beq
\vec{\rs} = \vec{\ri} - \frac{1}{|\vec{\ri}|^2} \vec{\ri}.
\label{eq:onelens}
\eeq
The above expression implicitly assumes that the lens is at the origin, and all the lengths have been normalized to the Einstein radius
corresponding to its mass (or equivalently, the lens mass has been assumed to be unity).

Let us consider the general case where we have $N$-point lenses, at
$\vec{r}_k=(x_k, y_k)$ with mass $M_k$, $k=1, \cdot\cdot\cdot, N$. We normalize
all the lengths with the Einstein radius corresponding to the total
mass, $M=\sum_{k=1}^N M_k$. The generalized lens equation then reads
\beq
\vec{\rs} = \vec{\ri} - \sum_{k=1}^{N} m_k \frac{\vec{\ri}-\vec{r}_k}{|(\vec{\ri}-\vec{r}_k)|^2},~~ m_k=\frac{M_k}{M},
\label{eq:nlens}
\eeq
where $\sum_{k=1}^N m_k =1$. If there is only one lens ($m_1=1$) and the lens is at the origin, then Equation (\ref{eq:nlens}) reverts to  the single lens equation (\ref{eq:onelens}).

Two-dimensional vectors and complex numbers are closely related, \cite{Wit90} first
demonstrated that the above equation can be cast 
into a complex form by direct substitutions of the vectors by complex numbers:
\beq
\zs = z - \sum_{k=1}^{N} m_k \frac{z-z_k}{(z-z_k)(\bar{z}-\bar{z}_k)}
= z - \sum_{k=1}^{N} \frac{m_k}{\bar{z}-\bar{z}_k},
\label{eq:complex}
\eeq
where $z=x + y\,i$, $z_k=x_k + y_k\,i$, and $\zs=\xs+ \ys\,i$ ($i$ is the imaginary unit).

We can take the conjugate of Equation (\ref{eq:complex}) and obtain an
expression for $\bar{z}$. Substituting this back into
Equation (\ref{eq:complex}), we obtain a complex polynomial of degree $N^2+1$.
This immediately shows that 1) even a binary lens equation cannot be solved analytically since it is a fifth-order polynomial,
and 2) the maximum number of images cannot exceed $N^2+1$. In fact, the
upper limit is $5(N-1)$ (see \sect\ref{sec:math}), which indicates that most
solutions of the complex polynomial are not real images of the lens equation.

The magnification is related to the determinant of the Jacobian of the mapping from the source plane to
the lens plane: $(\xs, \ys) \rightarrow (x,y)$. In the complex form,
this is (\citealt{Wit90}):
\beq
\mu = \frac{1}{J}, ~~~ J=\frac{\partial(\xs, \ys)}{\partial(x, y)} = 1 - \frac{\partial \zs}{\partial \bar{z}}
\overline{\frac{\partial \zs}{\partial \bar{z}}}.
\eeq
Notice that the Jacobian can be equal to zero implying
a (point) source will be infinitely magnified. The image positions satisfying $J=0$ form continuous
{\it critical curves}, which are mapped into {\it caustics} in the source plane. Of
course, stars are not point-like, they have finite sizes. The finite
source size of a star smooths out the singularity. As a result, the magnification remains finite.

\begin{figure}[!ht]
\hspace{-0.5cm}
\includegraphics[width=.5\textwidth]{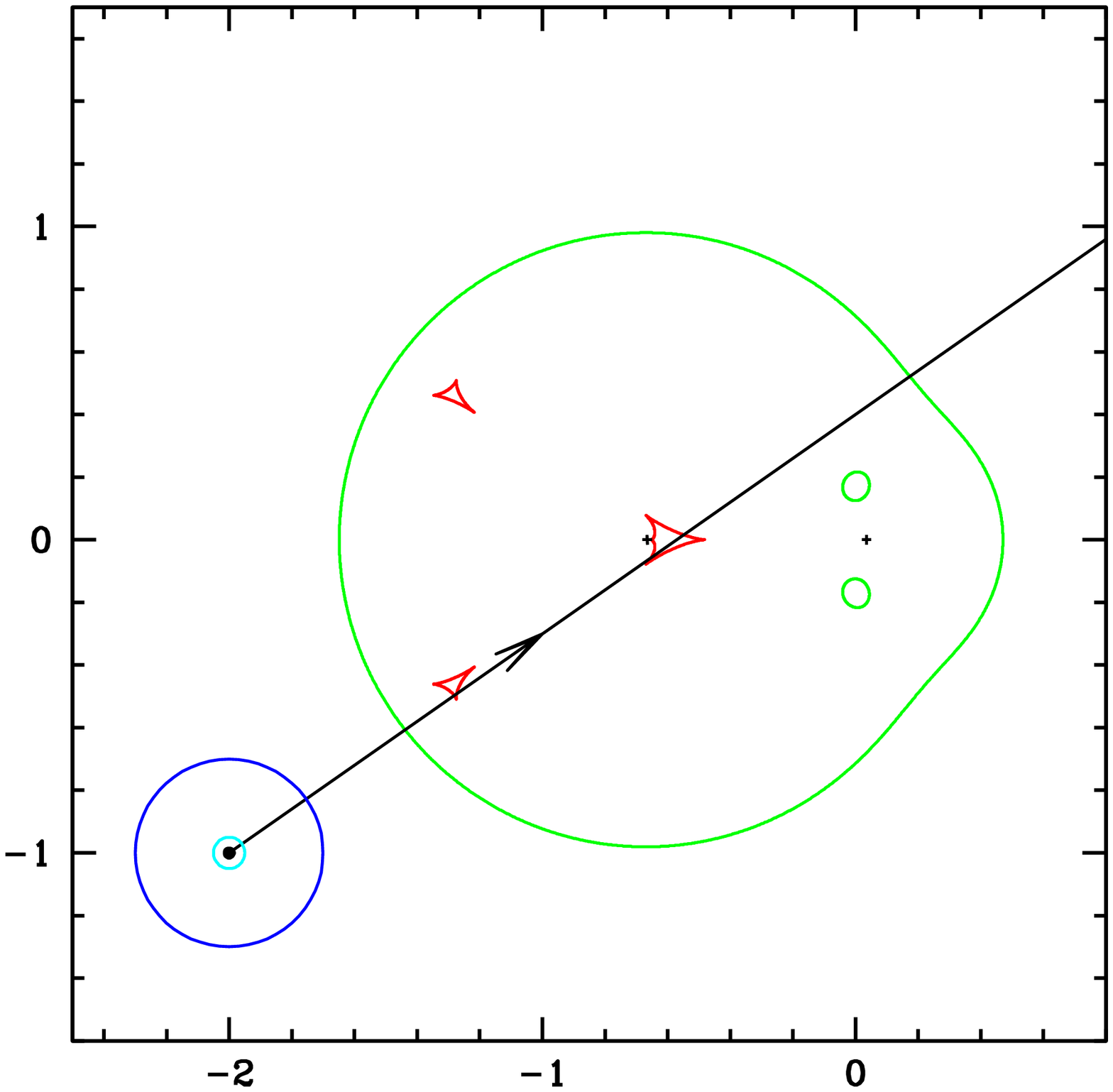} \hfil
\includegraphics[width=.5\textwidth]{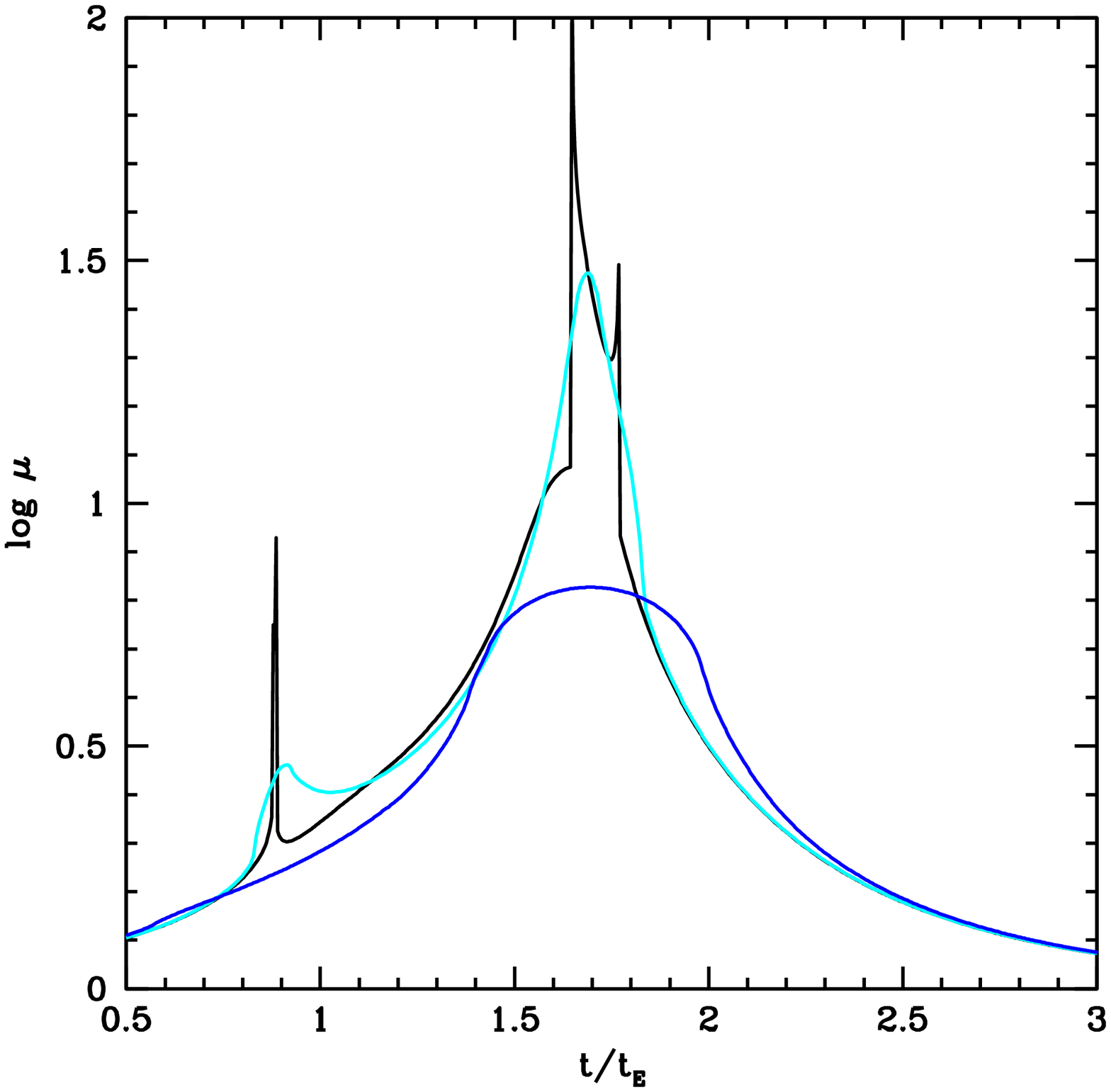}
\vspace{-0.5cm}
\caption{
{\it Left panel}: Caustics (red curves) and critical curves (green curves) for a binary lens. The
lenses (indicated by two `+' signs) are located at $z_1=(-0.665, 0)$ and $z_2=(0.035, 0)$ with mass $m_1=0.95$ and
$m_2=1-m_1=0.05$ respectively. The black line shows the trajectory
for three source sizes, $\rs/\rE=0, 0.05, 0.3$, indicated by the cyan
and blue circles and a dot (for a point source). The trajectory starts
at $(-2, -1)$ with a slope of 0.7. {\it Right panel}: Corresponding light curves for the three source sizes
along the trajectory in the left panel. The time is
normalized to the Einstein radius crossing time, $\tE$, and $t=0$
corresponds to the starting position. Notice that as the
source size increases, the lensing magnifications around the
peaks usually decrease. 
}
\label{fig:binary_lc}
\end{figure}

For $N$-point lenses, from the complex lens equation (\ref{eq:complex}),
we have
\beq
\frac{\partial\zs}{\partial \bar{z}} = \sum_{k=1}^{N}
\frac{m_k}{(\bar{z}-\bar{z}_k)^2}, ~~~
J = 1 - \Big|\sum_{k=1}^{N} \frac{m_k}{(\bar{z}-\bar{z}_k)^2}\Big|^2.
\label{eq:Jacobian}
\eeq
It follows that the critical curves are given by
\beq
\Big|\sum_{k=1}^{N} \frac{m_k}{(\bar{z}-\bar{z}_k)^2}\Big|^2 = 1.
\eeq
The sum in the above equation must be on a unit circle, and the solution can be cast in a parametric form
\beq
\sum_{k=1}^{N} \frac{m_k}{(z-z_k)^2} = {\rm e}^{i \Phi},
\eeq
where $0 \le \Phi < 2 \pi$ is a parameter. The above equation is a complex polynomial of degree $2 N$ with
respect to $z$. For each $\Phi$, there are at most $2N$ distinct solutions. As we vary $\Phi$ continuously, the
solutions trace out at most $2 N$ continuous critical curves (critical
curves of different solutions can smoothly join with each other). In practice, we can solve the equation for one $\Phi$ value, and then use the
Newton-Raphson method to find the solutions for other values of $\Phi$.
For a point source, the complex polynomial can be easily
solved numerically (e.g. using the {\tt zroots} routine in
\citealt{Pre92}, see also \citealt{Sko12}). However, for a source with finite size, the existence of
singularities makes the integration time-consuming (see \sect\ref{sec:future}).
The right panel in Fig. \ref{fig:binary_lc} shows the light
curves for three source sizes along the trajectory indicated by the
straight black line. As the source size increases, the lensing
magnifications around the peaks usually decrease and become broader. Furthermore,
the number of peaks may differ for different source sizes.


\subsection{Optical depth and event rates} \label{sec:tau}

So far we have derived the lens equation and light curve for
microlensing by a single star. In reality, hundreds of millions of
stars are monitored, and $\approx 2000$ unique microlensing events
are discovered each year (OGLE-IV alone identifies about 1500
microlensing events each year). 
Clearly we need some statistical quantities to describe microlensing experiments.
For this, we need two key concepts: optical depth and event rate.

\subsubsection{Optical depth}

The optical depth (lensing probability) is the probability that a given source 
falls into the Einstein radius of any lensing star along the line of sight. Thus the optical depth can be expressed as
\begin{equation}
\tau = \int_0^\Ds n(\Dd) \left( \pi \rE^2\right) d\Dd,
\end{equation}
which is an integral of the product of the number density of lenses, the lensing 
cross-section ($=\pi \rE^2$) and the differential path ($d\Dd$). 

Alternatively, the optical depth can be viewed as the fraction of sky covered by the angular areas of all the
lenses, which yields another expression
\begin{equation}
\tau = \frac{1}{4\pi} \int_0^\Ds \left[n(\Dd) 4\pi \Dd^2 d\Dd\right] \left( \pi \thetaE^2\right),
\end{equation}
where the term in the [~] gives the numbers of lenses in a spherical
shell with radius $\Dd$ to $\Dd+d\Dd$, $\pi\thetaE^2$ is the angular
area covered by a single lens, and the term in the denominator
is the total solid angle over all the sky ($4 \pi$).

If all the lenses have the same mass $M$, then $n(\Dd) = \rho(\Dd)/M$, $\pi
\rE^2 \propto M$, and the lens mass drops out in $n(\Dd) \pi \rE^2$.
Therefore the optical depth depends on the total mass density along the line
of sight, but not on the mass function.

Let us consider a simple model where the density is constant
along the line of sight, $\rho(\Dd) = \rho_0$. Integrating along the line of sight one finds
\beq
\tau = \frac{2 \pi G \rho_0}{3c^2} \Ds^2 = \frac{1}{2 c^2} \frac{G \rho_0 4\pi
  \Ds^3/3}{\Ds}
= \frac{1}{2 c^2} \frac{G M(<\Ds)}{\Ds} = \frac{V^2}{2 c^2},
\eeq
where $M(<\Ds)$ is the total mass enclosed within the sphere of radius
$\Ds$ and the circular velocity is given by $V^2 = GM(<\Ds)/\Ds$.

For the Milky Way, $V\approx 220\kms$, and $\tau \approx 2.6\times
10^{-7}$. The low optical depth means millions of stars have to be
monitored to have a realistic yield of microlensing events, and thus one
needs to observe dense stellar fields, which in turn means accurate
crowded field photometry is essential (see \citealt{Ala98, Ala00,
  Woz00, Woz08}, and references therein).

\subsection{Event rate} \label{sec:eventRate}

The optical depth indicates the probability of a given star that is within
the Einstein radii of the lenses {\it at any given instant}. As such, the optical
depth is a static concept. We are obviously interested in knowing the
event rate (a dynamic concept), i.e.,
the number of (new) microlensing events per unit time for a given number of
monitored stars, $N_\star$.

To calculate the event rate, it is easier to imagine the lenses are moving
in a background of static stellar sources. For simplicity, let us assume all the lenses move with the same velocity
of $\vt$. The new area swept out by each lens in the time interval $dt$ is equal to the product of the diameter of the Einstein
ring and the distance traveled  $\vt\,dt$, $dA= 2 \rE \times \vt dt = 2 \rE^2 dt /\tE$.
The probability of a source becoming a new microlensing event is given by
\beq
d\tau = \int_0^{\Ds} n(\Dd) dA d\Dd 
   =\int_0^{\Ds} n(\Dd) \left(\frac{2 \rE^2}{\tE}\right) dt d\Dd 
\eeq
The total number of new events is $N_\star d\tau$, and thus the event rate is given by
\beq
\Gamma = \frac{N_\star\,d\tau}{dt}  
   =N_\star \int_0^{\Ds} n(\Dd) \left( \frac{2}{\pi \tE} \cdot {\pi \rE^2}\right)\, d\Dd 
=\frac{2N_\star}{\pi}\int \frac{d\tau}{\tE}. 
\eeq
If, for simplicity, we assume that all the Einstein radius crossing times are
identical, then we have
\beq
\Gamma \approx \frac{2 N_\star}{\pi} \frac{\tau}{\tE}.
\label{eq:gamma}
\eeq
We make several remarks about the event rate:
\begin{enumerate}
\item[(1)] If we take $\tE=19 \day$ (roughly equal to the median of the observed timescales), then we
have
\beq
\Gamma \approx \frac{2 N_\star}{\pi} \frac{\tau}{\tE}
= 1200 \yr^{-1} \frac{N_\star}{10^8}\frac{\tau}{10^{-6}} \left(\frac{\tE}{19 \day}\right)^{-1}.
\eeq
For OGLE-III, about $2\times 10^8$ stars are monitored,
so the total number of lensing events we expect per year is
$\Gamma \sim 2400$ if $\tau \sim 10^{-6}$, which is within a factor of four of the observed rate (indicating the detection efficiency may be of the order of 30\%).
\item[(2)] While the optical depth does not depend on the mass function, the
  event rate does because of $\tE (\propto M^{1/2})$ in the denominator
  of Equation (\ref{eq:gamma}). The timescale distribution can
  thus be used to probe the kinematics and mass function of lenses in the Milky Way.
\item[(3)] The lenses and sources have velocity distributions; one must account for them
when realistic event rates are needed. Furthermore, the source distance
is unknown, and so in general we need to average over the source distance (for example calculations, see \citealt{Gri91, KP94}).
\end{enumerate}

\section{Applications of Gravitational Microlensing} \label{sec:apps}

As we can see, the theory of gravitational microlensing is relatively
simple.  When \cite{Pac86} first proposed microlensing, the astronomical community was
very skeptical whether microlensing events could be differentiated
from other types of variable stars (for an example of skepticism, see
\citealt{Mao08b}). As proven again and again in the field, such 
pessimism is often unwarranted: with the rapid increase in the
discovery rate of microlensing events (from a few events in the early
years to about 2000 events annually), even small probability events can be discovered.
The best example may be the seemingly crazy idea of
terrestrial parallax proposed by \cite{Har95} and \cite{Hol96}, which
has subsequently been observed by
\cite{Gou09}. Since the discovery of the first microlensing events in 1993, more than 10,000 events
(mostly in real-time) have been discovered. In the process,
tens of TBytes of data have been accumulated. This tremendous database
is a treasure trove for exploring
diverse astrophysical applications.

\subsection{MACHOs in the Galactic halo?}

Microlensing was originally proposed to detect MACHOs in the Galactic
halo \citep{Pac86}. Earlier results by the MACHO collaboration suggest
that a substantial fraction ($\approx 20\%$) of the halo may be in MACHOs 
\citep{Alc00} based on 13-17 events toward the Large Magellanic Cloud (LMC) in 5.7 years of data.
The number of events quoted here varies somewhat depending on the selection
criteria, which is a debated point (see, e.g., \citealt{Bel04, Ben05,
  Gri05, Eva07}).

In any case, the result turns out to be controversial, as these numbers are not
confirmed by the EROS \citep{Tis07} or OGLE collaborations \citep{Wyr11}. A recent analysis of the OGLE
data toward the LMC concluded only $\la 2\%$ of the halo could be in 
MACHOs \citep{Wyr11} - all events can be explained by lensing by normal stars. In
fact, at least two of the MACHO events turn out to be by stars in the thick disk of
the Milky Way \citep{Dra04, Kal06}. The lack of MACHOs is entirely consistent with the baryonic content
determined from many other astrophysical observations (e.g. from
microwave background observations, \citealt{Kom11}).

\subsection{Galactic Structure}

Microlensing data offer a number of ways to study the Galactic structure,
which is still somewhat under-explored, and more work is needed.

\subsubsection{Color-Magnitude Diagrams} 

Microlensing surveys yield many high-quality color-magnitude
diagrams of stellar populations over hundreds of square degrees. Such data
contain much information about stellar populations in the
bulge. Despite of some promising earlier attempts (e.g. \citealt{Ng96}),
not much work has been done since due to the difficulty of disentangling
the dust extinction, spatial distributions and complexities in the
stellar populations (see the next subsections).
It is nevertheless worthwhile to perform further analyses in this area.

\subsubsection{Red Clump Giants As Standard Candles}

Red clump giants are metal-rich, core-helium burning horizontal branch stars. They have
approximately constant luminosity (standard candles) 
\citep{Pac98} and relatively little dependence on the
metallicities, especially in the $I$-band (see \citealt{Uda98a, Zha01, NU11}, and references
therein). As such, they can be used to determine the
distances to the Galactic center and the Magellanic Clouds
\citep{Uda98b}. This complements the determination of
other standard candles such as RR Lyrae stars and Cepheids
\citep{Gro08}. The distance to the Galactic center determined with
these methods is typically very close to 8\,kpc, with a combined systematic and statistical error of about 5\%.

The observed width of red clump giants in a given line-of-sight toward the Galactic center is
larger than their intrinsic scatters ($\approx 0.2$\,mag) because of the
finite depth of the Galactic bar. A careful analysis of the counts of
red clump giants can be used to determine the geometric parameters of the bar. For
example, for 44 fields from the OGLE-II data, 
the three axial lengths are found to be close to
10:3.5:2.6 with a bar angle of 24-27$^\circ$
(\citealt{Rat07a}), largely consistent with the earlier study by \cite{Sta97}.
Extra care, however, needs to be taken for fields off the plane due to
the presence of two red clump populations \citep{Nat10}, which suggests
an X-shaped structure in the Milky Way center (see also \citealt{McW10, Sai11}).

We note that the OGLE-III and OGLE-IV campaigns cover much larger areas on the sky,
which can be used to constrain not only the bar but perhaps also spiral arm
structures. Investigations taking this approach are already under way
(Nataf et al. 2012, in preparation).

\subsubsection{Extinction Maps}

Red clump giants can also be used to infer extinctions toward the
Galactic Center. The first application of this method to the OGLE-I data was performed
by \cite{Sta96} and later to the OGLE-II data by \cite{Sum04}. The
extinction maps are approximately consistent with those obtained with
other methods (e.g. \citealt{Sch98}). One interesting conclusion from
these studies is that the extinction law is almost always anomalous toward the Galactic center \citep{Uda03}.

A very recent application of the method to the VISTA Variables in the Via Lactea (VVV) survey
can be found in \cite{Gon12}. 

\subsubsection{Proper Motions}

Nearly two decades of time series of observations for
many different bulge fields offer a way to determine stellar
proper motions. This has been done for the OGLE-II
catalog for about five million stars with an accuracy of $\sim {\rm
mas\,yr^{-1}}$ (see \citealt{Sum04b} and \citealt{Rat08} for a study of
selected high-proper motion stars). 

One particularly interesting issue concerns the anisotropy in the bulge
kinematics. The proper motion 
dispersions in the longitudinal and latitudinal directions have ratios
close to $\approx 0.9$; these values are similar to those independently measured with the Hubble
Space Telescope in a number of fields but for a smaller
number of stars \citep{Koz06, Cla08}. A comparison with theoretical
predictions seems to indicate that the
theoretical models are too anisotropic \citep{Rat07b}.This effect is also
seen in more recent modelling with the Schwarzschild method
\citep{Wan12}. The reason for this discrepancy is not completely
clear. One issue worth mentioning is that bulge observations
are often contaminated by disk stars, a point emphasized by \cite{Vie07}.

\subsubsection{Optical Depths and Galactic bar}

In \sect\ref{sec:tau}, we showed that the optical depth provides an
independent probe of the density distribution in the bulge. 
Earlier determinations of the optical depth in the Baade Window
have provided independent evidence that the Galactic
bulge contains a bar \citep{KP94, Zha96}.
While we have discovered about ten thousand microlensing
events over the last two decades, only a small fraction has been used
for statistical analyses of optical depths.

More recent determinations of microlensing optical depths  have been
performed by \cite{Ham06} using 120 red clump giants from the EROS collaboration, 60
red clump giants from the MACHO collaboration \citep{Pow05},
28 events for the MOA collaboration \citep{Sum03} and 32 high signal-to-noise ratio
events for the OGLE collaboration \citep{Sum06}. The measured
optical depths are largely consistent with theoretical predictions
(e.g., \citealt{Woo95, Ryu08, Ker09}).

The reason that the measurements have only been performed for red clump
giants is that they are bright and are supposed to suffer less from the
blending in the crowded bulge field (in fact, this is not rigorously
true, see \citealt{Sum06}). If we can overcome this and the issue of human resources (see the next subsection), combined with the number counts of red clump
giants, we can in principle constrain the bulge
density distribution much better using non-parametric models.
For example, we can use non-parametric models rather than the simple parametric models often used in the literature (still
based on very poor-resolution [$\approx 7^\circ$] maps from DIRBE on COBE, \citealt{Dwe95}).

\subsubsection{Timescale Distributions and Mass Functions}

As we have seen in \sect\ref{sec:eventRate}, the timescale distribution of
event rates carries information about the mass function of lenses. An
early study by \cite{Han95} of 51 events concluded that for a
power-law mass function, $n(M)dM \sim M^{-\beta}\,dM$, the preferred
slope $\beta \approx 2.1$, 
with a lower mass cutoff of about $0.04 M_\odot$. Their slope is
close to the Salpeter value ($\beta=2.35$).
The study by \cite{Zha96} also concluded that the data are 
inconsistent with a large population of brown dwarfs in the bulge.
A more recent analysis by \cite{Cal08} modeled  the mass function and 
found $\beta \approx 1.7\pm 0.5$. The large error
bar is again due to the small number of events used ($\sim 100$). 

The current data base
contains roughly two orders of magnitude more events - it would be very
interesting to explore what we can learn with the full data set we now
possess. However, there are at least two difficulties that will need to overcome. One is human resources: extensive Monte Carlo simulations must be performed
to determine the completeness of surveys - this will be a time-consuming
exercise. Secondly, the effect of blending of background stars needs to be taken into account (see \sect\ref{sec:standard}). This is
not so straightforward since the fields have different degrees of
crowding \citep{Smi07}. Nevertheless, it will be valuable to
perform an analysis of the optical depth maps and timescale
distributions using all the data set to understand the density
map and mass functions of the bulge and mass of the disk. Notice that
the microlensing sample is largely a  {\it mass}-selected sample, rather
than a light-selected sample as in most other studies.

This exercise may be particularly interesting in light of evidence for systematic
variations in the initial mass functions in elliptical galaxies from both
dynamics \citep{Cap12} and strong gravitational lensing \citep{Tre10}.
A careful study of the mass functions
in the bulge and disk from microlensing offers an important independent check on
these conclusions by investigating whether the Galactic bulge follows similar trends.

\subsection{Stellar atmosphere and bulge formation}

 High-magnification events offer great targets-of-opportunity
to obtain spectra with high signal-to-noise ratio to study
stellar atmospheres in bulge stars (\citealt{Len96}). More recent
observations using 8m class telescopes allow determinations of a
number of stellar parameters such as surface gravity, metallicity and ages
(e.g. \citealt{Len96, Thu06, Coh08}). 

The most recent studies by \cite{Bens10, Bens11} of 38 stars 
indicate that for dwarf 
stars, there may be two populations of stars. The metal-poor population
seems to be predominantly old (age $\sim$ 12 Gyr) while the metal-rich population appears to
have a bi-modal distribution: one is old ($\sim$ 12 Gyr) while the other one is
intermediate age (3-4 Gyr). The existence of intermediate age stars seems to be in
conflict with broad-band color and spectroscopy of giants
(e.g. \citealt{Zoc08}). One way to resolve this conflict is that the age estimate
may be incorrect due to enhanced helium enrichment \citep{Nat11}.
While more microlensed stars are desirable, 
it demonstrates the power of microlensing as a natural
telescope, similar to clusters of galaxies as a natural telescope to
study very high-redshift galaxies.

The surface brightness of a star is not uniform, instead its limb usually
appears darker (``limb-darkening''). Limb-darkening
profiles have been measured for the Sun and a few other stars. 
The sharp magnification gradient in high-magnification or
caustic-crossing events allows us to study the limb-darkening profile
as the source moves across the line of sight
or caustics. Notice that limb-darkening profiles can not only be studied in broad-band
but also in spectral lines such as H$\alpha$ if one has time-resolved
spectra during microlensing \citep{Thu06}.

Approximately 10 G and K giants had their limb-darkening
profiles measured with microlensing (\citealt{Cas06, Thu06, Zub11}).
These studies suggest that ``the classical laws are too restrictive to
fit well the microlensing observations'' \citep{Cas06}, and radiative
transfer models will need to be improved.

\subsection{Mass Determinations of Isolated Stars: from Brown Dwarfs to Stellar Mass Black Holes}

As shown in 
\sect\ref{sec:standard}, for standard microlensing events, the
lens mass cannot be determined uniquely. However, for exotic events,
partial or complete removal of this degeneracy is possible (see
equation \ref{eq:mass}). Microlensing thus provides an important new
 way to determine the mass of isolated stars.

All stellar black hole candidates in the Milky Way are in binary systems
and have been discovered through X-ray emissions (see \citealt{Rem06}
for a review). Their masses range from $5 M_\odot$ to $30 M_\odot$.  Microlensing provides an independent way to study isolated black holes. The principle is very
simple: Typical microlenses have masses of about $0.3 M_\odot$, black
holes are $\ga 10$ more massive, and so
events due to stellar mass black holes should be a factor of a
few longer. These events thus have a much larger chance of exhibiting
parallax signatures, which can be used to determine the lens as a function
of distance. Combined with a mass density and kinematic models of the Milky Way, we can
constrain the lens mass. If it is higher than a few solar
masses and it is dark (as can be inferred from the light curve from the
blending parameter $\Fs$, see \sect\ref{sec:standard}),
then it is a potential stellar mass black hole candidate.

Half a dozen or so such candidates have been identified by
(\citealt{Mao02, Ben02}, see also \citealt{Ago02, Poi05}). A search for the
X-ray emission in one of the candidates, MACHO-96-BLG-5, yielded only an upper limit
\citep{Mae05, Nuc06}, consistent with the system being a truly isolated
black hole.

An ambitious 3-year HST survey (192 orbits) is under way  by a team
led by K. Sahu to detect microlensing events caused by non-luminous
isolated black holes and other stellar remnants. It will be very
interesting to see the results from this survey \citep{Sah12}.

Other direct mass measurements have also been performed,
especially for highly magnified microlensing events. The determined masses 
range from brown dwarf candidates \citep{Smi03b, Gou09, Hwa10} to more
normal stars (see, e.g., \citealt{Bat09} and references therein).

\subsection{Extrasolar Planets} \label{sec:planets}

Undoubtedly the highlight of gravitational microlensing in the last
decade has been the discovery of extrasolar planets. At the time of writing,
about 20 microlensing extrasolar planets
have been discovered, including 16 published \citep{Bon04, Uda05,
  Bea06, Gou06, Ben08, Gau08, Don09b, Jan10, Sum10, Miy11, Bat11, Mur11,
  Yee12, Bac12, Ben12}. Of these, 11 are high-magnification events,
demonstrating they are excellent candidates for hunting planets, as
pointed out by \citep{GS98, Rat02}. While this is a small
fraction of the 800 extrasolar planets discovered so
far\footnote{http://exoplanet.eu/}, they occupy a distinct part of the
parameter space which would be difficult to access with
other methods (see below).

The method itself was proposed more than 20 years ago \citep{MP91,
  GL92}.  In the abstract of \cite{MP91}, it was optimistically claimed that
``A massive search for microlensing of the Galactic bulge stars may lead
to a discovery of the first extrasolar planetary systems.'' 
Paczy\'nski, however,
mentioned the idea as ``science fiction'' at the 1991 Hamburg
gravitational lensing conference. In reality, it took more than a
 decade of microlensing observations 
for the first convincing extrasolar planet to be found
 \citep{Bon04}, with heroic efforts in between. Much of the theory and
 observations have been reviewed by \cite{Rat06} and \cite{Gau10}; we
 refer the readers to those papers for further details.

\begin{figure}[!ht]
\centering
\vspace{-3mm}
\includegraphics[angle=-90,width=.8\textwidth]{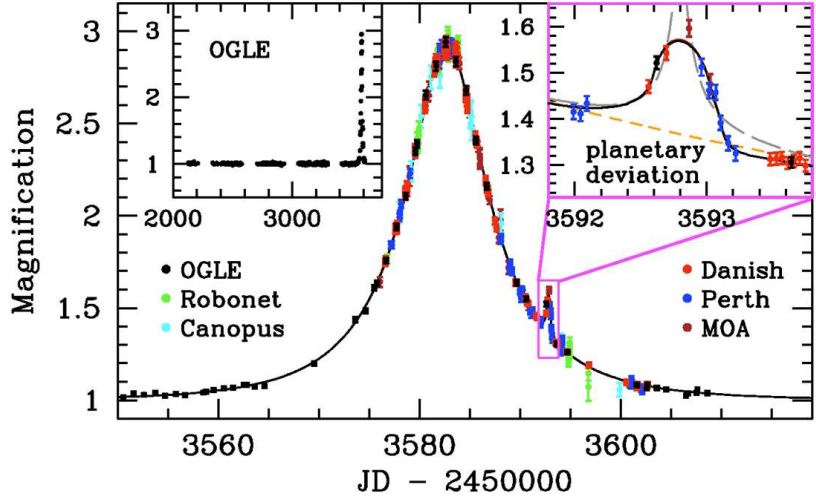}
\caption{A super-Earth ($\approx 5.5 M_{\earth}$) discovered by
  microlensing. Different symbols indicate data from different
  observatories; the solid line shows the best-fit model(s).
The two insets show the OGLE data alone and the planetary bump lasting
for about a day. Reprinted by permission from Macmillan Publishers Ltd: Nature 439:
437-440, \copyright2006.
}
\label{fig:beau}
\end{figure}

Fig. \ref{fig:beau} shows an example of an extrasolar planet discovered by
microlensing. The extrasolar planet has a mass of around $5.5 M_{\earth}$,
manifested as a secondary bump on the declining wing of the light curve,
lasting for about one day. This illustrates that to find an
extrasolar planet, the dense sampling of light curves plays a 
critical role.

\begin{figure}[!ht]
\centering
\vspace{-3mm}
\includegraphics[width=.7\textwidth]{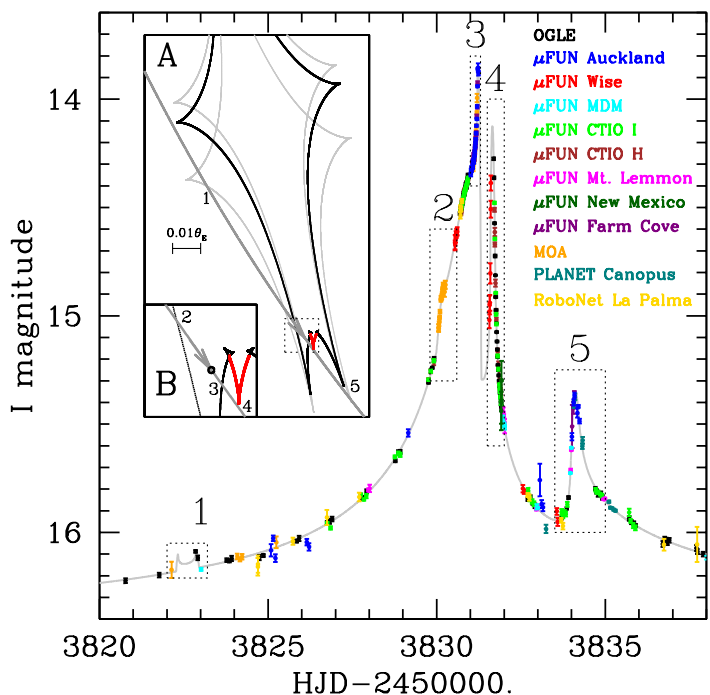}
\caption{The first two-planet system discovered by
  \cite{Gau08}. Different symbols are data taken from different
  telescopes (12 in total). The numbers indicate different features
 produced by caustic crossings and orbital motions. 
From Science, Volume 319, no. 5865, pp. 927-930. Reprinted with permission from AAAS.}
\label{fig:gaudi}
\end{figure}

The first two-planet system discovered by \cite{Gau08} through monitoring of high magnification
events is shown in Fig. \ref{fig:gaudi}. The light curve in this case is much more
dramatic due to the complex caustics involved (the top left
inset) and orbital motion in the system (see \citealt{Pen11} for detailed predictions). The mass and separations of the two
planets are very much like those for the Saturn and Jupiter in our solar
system. Such multiple planet systems constitute about
10\% of the extrasolar planets discovered through microlensing. Many
multiple planetary systems have been discovered in radial velocity and
transit surveys (e.g., \citealt{Wri09, Fab12} and references therein). In
radial velocity surveys, at least 28\% of known planetary systems appear to contain
multiple planets \citep{Wri09}. However, these two fractions cannot be
easily compared since the planets discovered are at very different
separations from the host stars. Furthermore
microlensing is only sensitive to planets within a narrow range of the
Einstein radius. A more detailed comparison is needed to address
this issue.

\begin{figure}[!ht]
\centering
\includegraphics[width=.9\textwidth]{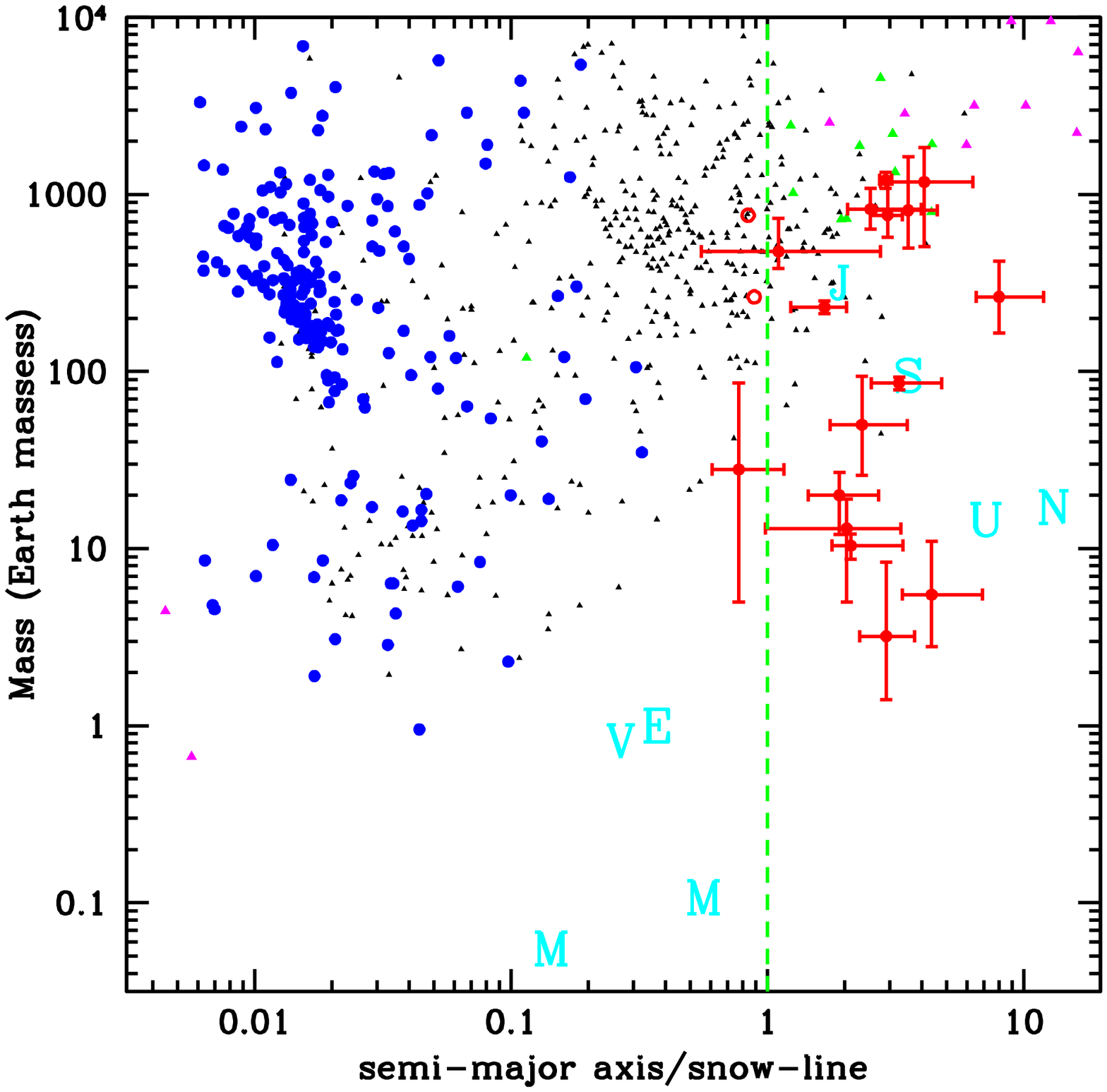}
\caption{Extrasolar planets in the plane of mass vs. separation (in units of the
  snow line, indicated by the vertical dashed green line). The snow line
  is taken to be at $\approx 2.7 M_{\rm L}/M_\odot$\, AU, where $M_{\rm
    L}$ is the lens mass (see Table \ref{tab:planets}). The red
filled circles with error bars indicate planets found by
microlensing. The open symbols show the degenerate
close-separation solutions for MOA-2007-BLG-400Lb and MOA-2011-BLG-293Lb
(see Table 1). The black triangles and blue squares indicate the planets discovered by
radial velocities and transits, respectively. The magenta and green triangles
indicate the planets detected via direct imaging and timing, respectively. The
non-microlensing exoplanet data were taken from The Extrasolar Planets
Encyclopaedia (http://exoplanet.eu/). The planets in our solar system are
indicated with initial letters (in cyan).
}
\label{fig:planets}
\end{figure}

Table \ref{tab:planets} lists all the published extrasolar planets
discovered by microlensing at the time of writing. This is an expanded
version of Table 3 in \cite{Miy11}. The microlensing planets occupy a
distinct part in the plane of  mass 
vs. separation normalized by the snow line. This is illustrated in Fig. \ref{fig:planets}. 
It is seen that microlensing planets reside mostly outside the snow line
where their equilibrium temperatures are low, about 100\,K (see also, e.g., Fig.\,3 in
\citealt{Bea08}).
Beyond the snow line, the density of solid particles in the
disk increases rather abruptly, the planetesimal cores can form
faster and the formation of gas giants becomes easier. The equilibrium
temperatures are very different from the hot planets found, for example, in radial
velocity searches. 

Analyses of microlensing extrasolar planets has already yielded very
interesting statistical results on planet frequency. Surprisingly,
six high-magnification events appear to form a well-defined ``sample'' \citep{Gou10}
even though the observations by the followup teams  were
triggered by somewhat chaotic human involvement.
The authors presented the first measurement of
the planet frequency beyond the snow line, for the planet-to-star
mass-ratio interval $-4.5 < \log q < -2$, corresponding to the range of
ice giants to gas giants. The frequency was found to follow
${d^2 N_{\rm pl}/ d \log q/ d \log s} = (0.36 \pm 0.15) {\rm dex}^{-2}$,
where $q$ is the mass ratio and $s$ is the separation. This is
consistent with the extrapolation of \cite{Cum08} to large separations.
Their study also implies a first estimate of 1/6 for the frequency of solar-like systems.

More recently, \cite{Cas12} reported
 a statistical analysis of microlensing data (gathered in 2002-07) and
 concluded that about 20\% stars host Jupiter-mass planets (between 0.3
 and 10  Jupiter masses) while cool Neptunes ($10-30 M_{\earth}$) and super-Earths
 ($5-10 M_{\earth}$) are even more common: their abundances per star are
 close to 60\%.

\begin{table}
\begin{center}
\begin{minipage}[]{\textwidth}
\caption[]{
Parameters of exoplanets discovered by microlensing. 
\label{tab:planets}
}
\end{minipage}
\tiny
\begin{tabular}{rccrcll}
\hline\noalign{\smallskip}
 Name & Host Star Mass   & Distance & Planet Mass  & Separation & Mass estimated 
by & References\\
\hline\noalign{\smallskip}
     & $M_{\rm L}(M_\odot)$ & $D_{\rm L}$(kpc) & $M_{\rm p}$ & $a$(AU)&& \\ 
\hline\noalign{\smallskip}
 OGLE-2003-BLG-235Lb & $ 0.63^{\ +0.07}_{\ -0.09}$ &$
 5.8^{\ +0.6}_{\ -0.7}$ & $ 2.6^{\ +0.8}_{\ -0.6} \ M_{\rm J}$ & $
 4.3^{\ +2.5}_{\ -0.8}$ & $\theta_{\rm E}$, LB & 1, 2\\
 OGLE-2005-BLG-071Lb & $ 0.46 \pm 0.04$ & $ 3.2 \pm 0.4$ & $ 3.8 \pm 0.4
 \ M_{\rm J}$ & $ 3.6 \pm 0.2$ & $\theta_{\rm E}, \pi_{\rm E}$, DL & 3, 4\\ 
 OGLE-2005-BLG-169Lb & $ 0.49^{\ +0.23}_{\ -0.29}$ & $ 2.7^{\ +1.6}_{\ -1.3}$ & $ 13^{\ +6}_{\ -8} \ M_\oplus$ & $
 2.7^{\ +1.7}_{\ -1.4}$ & $\theta_{\rm E}$, Bayesian & 5\\ 
 OGLE-2005-BLG-390Lb & $ 0.22^{\ +0.21}_{\ -0.11}$ & $ 6.6^{\ +1.0}_{\ -1.0}$ & $ 5.5^{\ +5.5}_{\ -2.7} \ M_\oplus$ & $
 2.6^{\ +1.5}_{\ -0.6}$ & $\theta_{\rm E}$, Bayesian &6\\ 
 OGLE-2006-BLG-109Lb & $ 0.51^{\ +0.05}_{\ -0.04}$ & $ 1.49 \pm 0.19$ &
 $ 231 \pm 19 \ M_\oplus$ & $ 2.3 \pm 0.5$ & $\theta_{\rm E}, \pi_{\rm
   E}$ &7\\ 
                  c & $ ^{}_{}$ & $ ^{}_{}$ & $ 86 \pm 7 \ M_\oplus$ &
                   $ 4.5^{\ +2.1}_{\ -1.0}$ & $\theta_{\rm E}, \pi_{\rm
                     E}$ &7\\
 OGLE-2007-BLG-368Lb & $ 0.64^{\ +0.21}_{\ -0.26}$ & $ 5.9^{\ +0.9}_{\ -1.4}$ & $ 20^{\ +7}_{\ -8} \ M_\oplus$ & $
 3.3^{\ +1.4}_{\ -0.8}$ & $\theta_{\rm E}$, Bayesian &8\\ 
 MOA-2007-BLG-192Lb & $ 0.084^{\ +0.015}_{\ -0.012}$ & $ 0.70^{\ +0.21}_{\ -0.12}$ & $ 3.2^{\ +5.2}_{\ -1.8} \ M_\oplus$ & $
 0.66^{\ +0.19}_{\ -0.14}$ & $\theta_{\rm E}, \pi_{\rm E}$ & 9\\ 
 MOA-2007-BLG-400Lb & $ 0.30^{\ +0.19}_{\ -0.12}$ & $ 5.8^{\ +0.6}_{\ -0.8}$ & $ 0.83^{\ +0.49}_{\ -0.31} \ M_{\rm J}$ & $
 0.72^{\ +0.38}_{\ -0.16}\ /\ 6.5^{\ +3.2}_{\ -1.2}$ & $\theta_{\rm E}$,
 Bayesian &10$^{a}$ \\ 
 MOA-2008-BLG-310Lb & $ 0.67 \pm 0.14$ & $ >6.0$ & $ 28 ^{\ +58}_{\ -23} \ M_\oplus$ & $ 1.4 ^{\ +0.7}_{\ -0.3}$ & $\theta_{\rm E}$, Bayesian & 11$^b$\\ 
 MOA-2009-BLG-319Lb & $ 0.38^{\ +0.34}_{\ -0.18}$ & $ 6.1^{\ +1.1}_{\ -1.2}$ & $ 50^{\ +44}_{\ -24} \ M_\oplus$ & $
 2.4^{\ +1.2}_{\ -0.6}$ & $\theta_{\rm E}$, Bayesian &12\\ 
 MOA-2009-BLG-387Lb & $ 0.19^{\ +0.30}_{\ -0.12}$ & $ 5.7^{\ +2.2}_{\ -2.2}$ & $ 2.6^{\ +4.2}_{\ -1.6} \ M_{\rm J}$ & $
 1.8^{\ +0.9}_{\ -0.7}$ & $\theta_{\rm E}$, Bayesian &13\\ 
 MOA-2009-BLG-266Lb & $ 0.56\pm 0.09$ & $ 3.04{\pm 0.33}$ & $10.4{\pm 1.7} M_\oplus$ & $
 3.2^{\ +1.9}_{\ -0.5}$ & $\theta_{\rm E}$, $\pi_{\rm E}$ &14$^{c}$\\ 
 MOA-2011-BLG-293Lb & $ 0.44^{\ +0.27}_{\ -0.17}$ & $7.15 \pm 0.65$
& $ 2.4^{\ +1.2}_{\ -0.6} M_{\rm J}$  & $1.0\pm 0.1$/$3.5\pm 0.5$ & $\theta_{\rm E}$, Bayesian &15$^d$\\ 
MOA-2010-BLG-477Lb & $ 0.67^{\ +0.33}_{\ -0.13}$ & $2.3 \pm 0.6$
& $ 1.5^{\ +0.8}_{\ -0.3} M_{\rm J}$  & $2^{+3}_{-1}$ & $\theta_{\rm E}$, Bayesian &16$^d$\\ 
  MOA-bin-1 & $ 0.75^{\ +0.33}_{\ -0.41}$ & $5.1^{+1.2}_{1.9}$
& $ 3.7\pm 2.1 M_{\rm J}$  & $8.3^{+4.5}_{-2.7}$ & $\theta_{\rm E}$, Bayesian &17\\ 
\hline
\noalign{\smallskip}
\end{tabular}
\end{center}
\tablenotes{a}{\textwidth}{
MOA-2007-BLG-400Lb has two solutions due to a strong close/wide model
degeneracy (see \sect\ref{sec:math}).}
\smallskip
\tablenotes{b}{\textwidth}{Details of the MOA-2008-BLG-310Lb parameters
  are discussed by \cite{Jan10} and \cite{Sum10}. The error bars take
  into account the degeneracy.}
{\vspace{-1.5mm}}
\tablenotes{c}{\textwidth}{90\% confidence limit.}
{\smallskip}
\tablenotes{d}{\textwidth}{The system has two solutions due to a close/wide
separation degeneracy (see \sect\ref{sec:math}), thus two
  separations are given from a Bayesian analysis assuming the lens is a
  main sequence star. The close separation is slightly favored.}
{\smallskip}
\tablecomments{\textwidth}{LB: lens brightness; DL: detection of the
lens.} 
{\smallskip}
\tablerefs{\textwidth}{
1. \cite{Bon04};
2. \cite{Ben06};
3. \cite{Uda05};
4. \cite{Don09a};
5. \cite{Gou06};
6. \cite{Bea06};
7. \cite{Gau08};
8. \cite{Sum10};
9. \cite{Ben08};
10. \cite{Don09b};
11. \cite{Jan10};
12. \cite{Miy11};
13. \cite{Bat11};
14. \cite{Mur11};
15. \cite{Yee12};
16. \cite{Bac12};
17. \cite{Ben12}.
}
\end{table}

Somewhat controversially, the study by \cite{Sum11} found a population
of free-floating Jupiters, with 1.8 Jupiters per star on average. These
planets are not bound to any parent stars, observationally they manifest
as very short-time scale events
(in contrast to very long events for stellar mass black hole candidates).
Whether such a high frequency of free-floating Jupiter-mass planets can be
produced in core accretion theory or gravitational instability theory is unclear.

\subsection{Numerical Methods in Modelling Binary and Multiple Light Curves}
\label{sec:num}

The modeling of light curves can be naturally divided into two parts:
The efficient calculation of light curves and finding the best-fit
models in a $\chi^2$ sense. We refer the readers to \cite{Ben10} for an
excellent discussion of this topic.

\subsubsection{Calculation of light curves}

For a single lens, the image magnifications and positions are
analytical. The  light curve calculation is straightforward, even
accounting for finite source size \citep{WM94}.

For a finite size source lensed by a binary or multiple system, its magnification
can be calculated in three different regimes with increasing
complexities. 
\begin{itemize}
\item When the source is far away from the caustics or any sharp
  magnification gradient, then the point source approximation can be
  used. In this case, the magnification can be readily found by solving
  the fifth-order or the tenth-order polynomials for binary or triple
  lens systems (e.g. using the {\tt zroots} routine in
\citealt{Pre92}; for a more efficient implementation avaliable publically see \citealt{Sko12}).
In practice, as the source moves along its trajectory,
  we can feed the image positions from the previous step as the initial
  guesses and use the Newton-Raphson (Simpson 1740) method to
  achieve rapid convergence to new solutions for the current
  source position. These can be used to deflate
  the polynomial to a lower-order one, which can then be solved much
  more easily.

  In the case of binary lensing, the deflated polynomial is
  usually either linear or quadratic and its solutions can then be found
  analytically. In this approach, we, at all times, keep all the five solutions to the
  polynomial (not all are true solutions of the lens equation).
  This efficient method was used in the first attempt to predict
  binary light curves \cite{Jar01}; it is similar to \cite{Sko12} in
  spirit. We note that for five-image configurations, the total magnification of positive-parity images
  must exceed that of negative-parity images by unity \citep{WM95}. This
  can be used as a test of numerical accuracy.
\item When the source is moderately (one diameter or so) far away from
  the caustics, the magnification can be efficiently calculated using
  the hexadecapole approximation proposed by \cite{Gou08}. It has the
  advantage of being very efficient and at the same time providing an
  estimate of the approximate accuracy (see also \citealt{Pej09}).
\item When the source is approximately within one diameter of the
  caustics, the magnification for a finite source size calculation becomes complex and
  time-consuming due to the presence of singularities.
  Many methods have been proposed, including (modified) rayshooting
  \citep{Rat02, Don06}, level contouring (based on Stokes theorem,
  \citealt{GG97, Dom07}) and grid  integration \citep{BR96, Ben10}. For detailed comparisons between these
  methods, see \cite{Ben10}. It is clear that there is still room for
  improvement. Furthermore, several independent modeling codes are
  desirable for cross-checks.
\end{itemize}

\subsubsection{Finding the global minimum}

Once a binary or multiple light curve can be efficiently calculated, it
still remains a highly non-trivial job to find the best-fit parameters,
especially in the presence of degeneracy (see
\sect\ref{sec:math}). Starting from an initial guess of the parameters, many
routines can be used to converge to a local minimum using, e.g., 
MINUIT \footnote{http://wwwasdoc.web.cern.ch/wwwasdoc/minuit/minmain.html} 
or GSL routines \footnote{http://www.gnu.org/software/gsl/}. 
More recently the Monte Carlo Markov Chain method has gained popularity
(e.g. \citealt{Don06, Ben10}).

However, to find the global minimum, often a grid of initial guesses are used.
This may be, however, impractical once the dimension of the parameter
space increases, especially for multiple lenses (for a nice discussion
of the parameters in binary lensing including parallax and orbital motion, see \citealt{Sko11}). This may be particularly
severe when the next-generation experiments come online; this area
deserves much more research in the future (see 
\citealt{Ben10} for the current state of affairs).

\subsection{Mathematics of Gravitational Microlensing}
\label{sec:math}

Gravitational lensing is mathematically rich and is related to
singularity (catastrophe) theory (see the monograph by \citealt{Pet01}).

In classical Keplerian potential theory, the two-body problem is analytically
tractable. In gravitational lensing, as we have shown in
\sect\ref{sec:binary}, it is not even possible to solve the binary lens
equation analytically. Curiously, for binary lenses, there is still an analytical theorem that
five-image configurations must have a magnification no smaller than
3 \citep{WM95,  Rhi97}. This was used to infer the presence of blending
in gravitational microlensing \citep{WM95}.

In binary lensing, there is a degeneracy found by \cite{Dom99}
between close and wide separation binaries. This was later explored in
much greater detail by \cite{An05}. Planetary and binary lens light
curves can also mimic each other \citep{Cho12}. It is unclear whether some of these
degeneracies can be generalized to multiple ($N \ge 3$) lenses.
For parallax events, there are also degeneracies \citep{Smi03a, Gou95}.

The number of critical curves for $N$-point lenses cannot exceed $2N$
(see \sect\ref{sec:binary}). The upper bound on the number of
images for $N$-point lenses is $5(N-1)$ \citep{Rhi03, Kha06}.
It is unclear whether these two linear scalings are related in a topological sense (see
\citealt{Rhi01, Rhi03, An06}). 

\section{Future of Gravitational Microlensing} \label{sec:future}

The future of gravitational microlensing is bright. In terms of
current experiments, OGLE-IV already has a field of view of 1.4 square
degrees and is readily
identifying about 1500 events in real-time per year. The rate can in principle
be increased by $\sim 40\%$ by analyzing their archives. 

The 1.8m MOA-II
telescope has a field of view of 2.2 square degrees and the MOA
collaboration issues about 600 microlensing alerts each year. 
The MOA collaboration uses difference image analysis photometry to issue alerts, which can identify faint source
stars that are undetectable when unmagnified. This is in contrast to the
OGLE, which to date has only been triggered by stars identified using their template
image. The MOA strategy appears to increase the alert rate by roughly a
factor of $\sim 1.4$ compared to OGLE. 

There has also been an influx of new telescopes
involved in microlensing, including the WISE Observatory which is
conducting an independent survey. In the future, more
telescopes may engage in microlensing campaigns, including the SKYMAPPER. 
Also, Chinese astronomers are discussing possible
monitoring from Antarctica (Dome-A) and Argentina where 2m class
telescopes have been proposed by the Chinese Academy of Sciences. The
superior seeing and continuous time coverage for a few months
per year at Dome-A may be an advantage for microlensing (although the seeing is
degraded due to the high air mass, \citealt{Wan09}). In fact, as the
first step, the first
of three 50\,cm telescopes has already been installed at Dome-A in early
2012 and will perform pilot surveys.

In terms of hunting for extrasolar planets, several White Papers 
  (\citealt{Gou07, Ben07, Dom08, Bea08}) set out strategies with ambitious
  milestones for the next fifteen years. 
\begin{itemize}
\item Current extrasolar discovery with microlensing is a mixture of
  surveys and intense followup observations. However, the boundary between surveys
  and followup teams is already starting to blur, with survey teams for some
  fraction of their time, engaging in very dense monitorings of some fields, for
  example, by both the MOA \citep{Sum11} and OGLE teams.
\item In the next five years, a network of three 1.6m telescopes will be
  built by Korean astronomers (Korean Microlensing Network,
  KMTNet). These telescopes will be sited in Chile, South Africa and Australia. Each is equipped
 with a 4 square degree field camera, and will be able to monitor 16
 square degrees and densely sample light curves
(every 10 minutes or so). This will in principle get rid of the cumbersome
division between survey and followup. As a result, the
selection functions will be much better specified and thus statistical
studies will be easier to perform. This is important as
the number of microlensing extrasolar planets will increase significantly.
\item A microlensing telescope in space in the next 10-15
  years has been proposed both in the US and Europe. WFIRST is the top
recommended space mission by the US Decadal
  Survey \citep{Bla10}. The recently funded Euclid mission by ESA may also have
  a microlensing component (for detailed simulations, see \citealt{Pen12}). Observing from space has substantial
  advantages: smaller and more stable PSFs and continuous time coverage
will allow us to search for lower-mass (Earth-mass)
  planets. Space missions will also allow us to uniquely determine the masses of many extrasolar planets. Combined with the stellar transit mission {\it   Kepler}, 
  space microlensing experiment(s) will provide a complete census of Earth-mass (and lower-mass) planets at
  virtually all separations, including free-floating ones.
\end{itemize}

The theory of microlensing is well understood, although computationally
there are still some challenging issues (see \sect\ref{sec:num}). For example, 
it is still time-consuming to calculate the light curves for
finite-size sources since we need to integrate over the
singularities of caustics. This is particularly important for the discovery of extrasolar
planets when a source transits the small caustics induced by the planet(s).
The problem becomes even worse for multiple planets
(\citealt{Gau08}). How do we efficiently search the high
dimensional parameter space? Are there hidden multiple
planetary light curves in the database that are not yet identified due to their complex shapes?

With a very healthy interplay between theory and observations,
upgraded/new surveys in the near term and space satellites on the horizon,
microlensing can expect another exciting decade in the future.

\normalem
\begin{acknowledgements}
I thank Drs. Liang Cao, Lijun Gou, Andy Gould and Richard Long for
many helpful discussions and criticisms on the review.
\end{acknowledgements}

\end{document}